\documentstyle[preprint,tighten,prc,aps,epsfig]{revtex}

\begin{document}

\title{Photoproduction of $\eta^\prime$-mesons from the Proton}
\author{A. Sibirtsev$^{1,2}$, Ch. Elster$^{2,1}$, S. Krewald$^{1}$,
and J.~Speth$^{1}$}

\address{
$^1$Institut f\"ur Kernphysik, Forschungszentrum J\"ulich,
D-52425 J\"ulich, Germany \\
$^2$Institute of Nuclear and Particle Physics,
Ohio University, Athens, OH 45701}

\date{\today}
\maketitle

\begin{abstract}
The presently available data for the reaction ${\gamma}p{\to}\eta^\prime{p}$ 
are analyzed in terms of a model in which the dominant production mechanism 
is the exchange of the vector mesons $\omega$ and $\rho$.  To describe the 
data at photon energies close to the production threshold we introduce a 
resonance contribution due to the well established $S_{11}(1535)$ resonance. 
Finally we study the contributions due to nucleon exchange to the $\eta^\prime$ 
photoproduction and find, that those contributions can be seen at
large angles in the differential cross section.
\end{abstract}

\maketitle

\section{Introduction}
The investigation of the production of $\eta^\prime$ mesons  in hadron 
and photon induced reactions provides  direct access to low energy QCD 
dynamics. The intrinsic properties of the $\eta^\prime$ meson are 
dominated by the gluonic degrees of freedom and continue to attract 
theoretical and experimental interest since the first observation of 
$\eta^\prime$ mesons~\cite{Kalbfleisch,Goldberg} in the reaction
$K^-p{\to}\Lambda\eta^\prime$  at LBL and BNL in 1964.
 
The mass of the $\eta^\prime$ meson is substantially larger than the ones of the
other mesons from the pseudoscalar meson nonet. Therefore, the $\eta^\prime$ meson 
could not be considered as  pure pseudoscalar Goldstone boson with a mass generated 
by spontaneous chiral symmetry breaking. However, the strong chiral $U_A(1)$ 
anomaly in QCD allows the  $\eta^\prime$ meson to gain additional mass by different 
mechanisms~\cite{Weinberg,Rejula,Hooft1,Hooft2,Witten,Veneziano1,Christos}. 
The large mass of the $\eta^\prime$ meson can be viewed as being generated
essentially by 
non-perturbative gluon dynamics and the $U_A(1)$ axial anomaly. 

Furthermore, an anomalous gluon not only generates a large  $\eta^\prime$ 
meson mass, but also affects the dynamics of the $\eta^\prime$-nucleon interaction. 
The flavor singlet Goldberger-Treiman relation   
correlates~\cite{Veneziano2,Shore1,Hatsuda,Efremov,Chao,Cheng}
the $\eta^\prime$-nucleon coupling constant, $g_{\eta^\prime{NN}}$,
and the flavor singlet axial charge of the nucleon,
$g_A^0$, and in the chiral limit~\cite{Shore2,Bass1} one has
\begin{equation}
g_A^0=\left.\sqrt{\frac{3}{2}}\right.\, \frac{F_0}{m_N} \, (g_{\eta^\prime{NN}}-g_{GNN}).
\label{GD}
\end{equation}
Here $m_N$ is the nucleon mass, $F_0{=}0.1$~GeV renormalizes the flavor
singlet decay constant~\cite{Shore1,Bass2}, and $g_{GNN}$ is the 
gluon-nucleon coupling~\cite{Shore1}. Moreover, $g_A^0$ can be decomposed in terms 
of the quark and gluon spin components of the proton~\cite{Altarelli,Carlitz}.

Unexpectedly, a very small value of $g_A^0{=}0.2{\div}0.35$ was extracted from 
EMC measurements~\cite{Ashman} of the polarized proton structure function. In a naive 
parton interpretation, i.e. neglecting the gluonic degrees of freedom by 
taking $g_{GNN}{=}0$, the EMC result indicates that the quark contribution
to the proton spin,
$\Delta\Sigma{=}\sqrt{3/2}\,g_{\eta^\prime{NN}}\,F_0/m_N$,  
is compatible with zero. This interpretation has supported speculations either 
about the validity of experimental results and data analysis
or the applicability of perturbative QCD.
Alternatively, the spin crisis can be  resolved by substituting a large
positive value, $g_{GNN}{\simeq}2.45$, and substantial cancellation between the
quark and gluon spin components. It was suggested~\cite{Chao,Ji}, that instead of 
the gluon component the contribution from the $\eta^\prime$ meson is canceled 
by the higher mass states, which are either radial excitations  of the 
$\eta^\prime$ meson or glueball states.

At present, the two component interpretation of the axial charge is generally 
accepted. However, both $g_{\eta^\prime{NN}}$ and $g_{GNN}$ coupling constants 
are not known individually. The origin of the proton spin and
its decomposition in terms of quark and gluon degrees of freedom still
remain controversial. However, it is believed that  the coupling constant
$g_{\eta^\prime{NN}}$ defines
up to large extent  the total spin of the nucleon 
carried by its constituents. 

It is important to note that the coupling constant  $g_{\eta^\prime{NN}}$ 
may experimentally be accessible 
in nuclear reactions involving the coupling between the $\eta^\prime$ meson and the 
nucleon. 

The reaction $pp{\to}pp\eta^\prime$ was intensively studied at  
COSY~\cite{Moskal1,Moskal2,Khoukaz,Moskal3} and SATURNE~\cite{Hibou,Kuhn,Balestra}.
Theoretical calculations~\cite{Sibirtsev1,Bernard,Baru,Nakayama}
allow  a description of the data entirely by the meson exchange currents. It was found 
that the contribution of the nucleon current to the  $\eta^\prime$ meson production in
proton-proton collisions is small. Thus the reaction $pp{\to}pp\eta^\prime$
is not well suited for the direct 
evaluation of the coupling constant $g_{\eta^\prime{NN}}$. Additional measurements of
differential cross sections 
and more detailed theoretical studies are necessary to determine
under what  conditions the nucleon exchange current may dominate
the meson exchange or may be at least partially isolated.

Experimental information~\cite{Landolt} about 
the reactions $\pi^-p{\to}\eta^\prime{n}$ and 
$\pi^+n{\to}\eta^\prime{p}$  is insensitive to the contribution from the
nucleon exchange, although in principle these reactions would be
 well suited for the determination
of the coupling constant $g_{\eta^\prime{NN}}$. Here the situation may be 
substantially improved with  new and  precise data for  
$\eta^\prime$ meson production at large angles, where the nucleon current
dominates.

The reaction $\gamma{p}{\to}\eta^\prime{p}$ was studied by a number of
collaborations, namely ABBHHM~\cite{ABBHHM},
AHHM~\cite{AHHM} and SAPHIR~\cite{SAPHIR1,SAPHIR2}. While
ABBHHM and  AHHM measured only the total cross section for the photoproduction 
of the $\eta^\prime$ meson
at photon energies $1.67{\le}E_\gamma{\le}5$~GeV, the SAPHIR collaboration at ELSA
studied the angular spectra of  the $\eta^\prime$ mesons produced at photon 
energies $0.9{\le}E_\gamma{\le}2.6$~GeV. In addition, the photoproduction of
$\eta^\prime$ mesons 
from the nucleon is presently experimentally  
investigated by the CLAS Collaboration at TJNAF~\cite{CLAS1}
and by the Crystal Barrel Collaboration at ELSA.

Currently experimental and 
theoretical~\cite{Zhang1,Zhang2,Li,Borasoy1,Borasoy2,Chiang}
studies of the reaction ${\gamma}p{\to}\eta^\prime{p}$ 
are motivated by the possibility to 
investigate  excited baryons coupled to the $\eta^\prime$ meson.
Assuming that the resonance production is the dominant contribution, 
an isobar analysis~\cite{SAPHIR1} of the SAPHIR data
shows that the $\eta^\prime$-photoproduction at photon energies
$0.9{\le}E_\gamma{\le}2.6$~GeV can be described by 
the coherent excitation of two resonances, $S_{11}(1897)$ and 
$P_{11}(1986)$. However, the presently available 
data~\cite{ABBHHM,AHHM,SAPHIR1,SAPHIR2} 
for the $\eta^\prime$-photoproduction are quite limited. 
Thus, an  evaluation of resonance properties from these data should
be considered with caution. It is also clear that the contribution from the
meson exchange current to the reaction $\gamma{p}{\to}\eta^\prime{p}$ 
can not be completely neglected, as was shown in the analysis of the
reaction $pp{\to}pp\eta^\prime$.

Here we investigate up to what extent present data for the reaction
${\gamma}p{\to}\eta^\prime{p}$ can be described by the
exchange of vector mesons in the $t$ channel and the 
nucleon exchange in the $s$ and $u$ channels. We do not consider 
contributions from baryonic resonances with masses above the production threshold. 
If possible resonances will significantly contribute 
to the production mechanism, 
then a discrepancy between our calculation and data should 
indicate this. 

In addition, we aim to investigate the conditions under which the nucleon exchange
current can be well controlled, and whether the reaction $\gamma{p}{\to}\eta^\prime{p}$ 
may be used to better determine the coupling constant 
$g_{\eta^\prime{NN}}$.

\section{Vector Meson exchange as production mechanism}
One of the well established results in particle physics is that for very 
high energies, i.e. $E_\gamma >$5~GeV, Regge trajectories provide the
dominant processes for peripheral reactions. At energies well below 5~GeV, 
but still clearly above the resonance region 
the photoproduction of mesons can be described  the $t$ channel meson exchanges.
In our case the photoproduction
of the $\eta^\prime$ meson
is dominated by the exchanges of the $\rho$ and the $\omega$ meson. The relevant 
Feynman diagrams are shown in Fig.~\ref{diagram1}. The four momenta of the
initial nucleon and photon and the final nucleon and $\eta^\prime$ meson
are denoted as $p_i$, $k$, $p_f$ and $q$, respectively.  

We use the Mandelstam invariants to define the reaction kinematics and
invariant reaction amplitudes, namely
\begin{eqnarray}
s&=&(p_i+k)^2=(p_f+q)^2=m_N^2+2m_NE_\gamma \nonumber \\
t&=&(p_i-p_f)^2=(k-q)^2 \nonumber \\
u&=&(k-p_f)^2=(p_i-q)^2=2m_N^2{+}m^2_{\eta^\prime}{-}t{-}s.
\end{eqnarray}
The relation between the four momentum transfer squared and the 
scattering polar angle $\theta$ is given as
\begin{equation}
t=m^2_{\eta^\prime} -2q^0k^0+2|{\bf q}||{\bf k}| \cos\theta.
\end{equation}
Thus, $\eta^\prime$ meson photoproduction at backward angles, 
$\cos\theta{=}-$1, corresponds to the maximal four momentum
transfer squared, -$t$, and increases with increasing the
photon energy $E_\gamma$. Moreover the maximal --$t$ corresponds 
to the minimal value of $|u|$. At the production threshold for  $\eta^\prime$
mesons
the squared invariant collision energy is given by 
$s{=}(m_N{+}m_{\eta^\prime})^2$, and the invariant  $t$   takes the form
\begin{equation}
t = -\frac{m_N m_{\eta^\prime}^2}{m_N + m_{\eta^\prime}}
=-0.45~{\mbox GeV^2}.
\label{eq:4}
\end{equation}
From Eq.~(\ref{eq:4}) it is obvious, that already at the  reaction threshold
 the $\eta^\prime$ meson photoproduction
probes large $|t|$. Thus, the $\rho$ and $\omega$  meson exchanges should be very
sensitive~\cite{Sibirtsev2} to the choice of the form factors at the nonlocal 
interaction vertices. The shaded area in Fig.~\ref{prim6} shows the range
of the four momenta transfer squared $t$ accessible in the
reaction $\gamma{N}{\to}\eta^{\prime}N$ 
at a given photon energy $E_\gamma$, or at a given squared invariant collision
energy $s$. The figure shows clearly that
for the energies  available at ELSA and TJNAF, namely $E_\gamma{<}2.5$~GeV, 
 the photoproduction of $\eta^\prime$- mesons probes large $|t|$ up to
$\simeq$-3~GeV$^2$.


Furthermore, it is important to notice that 
uncertainties in the  $t$-channel contributions due to unknown coupling 
constants and  form factors can be eliminated through data collected at energies 
slightly beyond the resonance region. Here the $t$-channel contributions 
dominate at the low four momentum transfer squared.  
We also note  
that at high energies, 
$E_\gamma{>}$5~GeV, the elementary particle exchange should be generalized 
to the correct expression for the exchange of a Regge trajectory with physical 
coupling constants comprising the 
residue \cite{Donnachie1,Donnachie2,Sibirtsev3}. Thus, the availability of
precision  data 
at photon energies $2{<}E_\gamma{<}5$~GeV  is very crucial for the theoretical 
analysis of $t$-channel exchanges.

The effective Lagrangian densities used for the evaluation of the vector 
meson ($V$) exchange amplitudes are given 
as~\cite{Levy,Olsson1,Olsson2,Benmerrouche1,Benmerrouche2,Sibirtsev4,Tsushima1}
\begin{eqnarray}
{\cal L}_{VNN} &=& g_V {\bar N} \gamma_\mu N V^\mu +
\frac{g_T}{2m_N} {\bar N} \sigma_{\mu\nu} N V^{\mu\nu} \\
{\cal L}_{V\eta^\prime\gamma} &=&  \frac{e g_{V\eta^\prime\gamma}}
{m_{\eta^\prime}} \varepsilon_{\mu\nu\alpha\beta} 
F^{\mu\nu} V^{\alpha\beta} \eta^\prime.
\label{lag1}
\end{eqnarray}
Here $m_{\eta^\prime}$ stands for the mass of the $\eta^\prime$ meson, $m_N$ for
the mass of the nucleon. The vector meson
field tensor is given by
$V_{\mu\nu}{=}\partial_\nu{V_\mu}{-}\partial_\mu{V_\nu}$, 
and $F^{\mu\nu}{=}\partial_\nu{A_\mu}{-}\partial_\mu{A_\nu}$ with $A^\mu$ 
being the photon field.
 
The  vector ($g_V$) and tensor ($g_T$) coupling constants at the $\omega{NN}$ and 
$\rho{NN}$ vertices are adopted from the Bonn potential model~\cite{Elster}. 
As in the NN potential model, we dress the vertices with 
monopole form factors,
\begin{equation}
F_{VNN}(t)= \frac{\Lambda_{VNN}^2-m_V^2}{\Lambda_{VNN}^2-t},
\label{ff1}
\end{equation}
with $m_V$ being the mass of the vector meson. We adopt the cut-off parameter
$\Lambda_{VNN}$=1.5~GeV from the Bonn model, as well as the coupling contants
$g_{\rho{NN}}$=3.9 and $g_{\omega{NN}}$=10.6. The ratio of the tensor to vector
coupling is 6.1 for the $\rho$ meson and zero for the $\omega$ meson.

The couplings for the $\rho\eta^\prime\gamma$ and 
$\omega\eta^\prime\gamma$ vertices can be evaluated from the 
partial decay widths of the reactions
$\eta^\prime{\to}\gamma\rho$ and $\eta^\prime{\to}\omega\rho$.
The two quantities are related by 
\begin{equation}
\Gamma_{\eta^\prime\to \gamma V} = \frac{e^2g^2_{V\eta^\prime\gamma}}{32\pi}
\frac{(m_{\eta^\prime}^2-m_V^2)^3}
{m_{\eta^\prime}^5},
\end{equation}
where $e^2{=}4\pi\alpha$, with $\alpha$ being the electromagnetic coupling.
With $\Gamma_{\rho\to\gamma\eta^\prime}{=}60{\pm}5$~keV and 
$\Gamma_{\omega\to\gamma\eta^\prime}{=}6.1{\pm}0.8$~keV we obtain 
$g_{\rho\eta^\prime\gamma}$=1.36, and $g_{\omega\eta^\prime\gamma}$=0.4.

The form factors at the $\rho\eta^\prime\gamma$ and 
$\omega\eta^\prime\gamma$ vertices are not fixed by the model. As  shown in
Fig.~\ref{prim6}(a), the $\eta^\prime$- meson photoproduction probes a large
region of $|t|$. 
Thus we expect that our calculations may be very sensitive to the 
choice of the unknown form factor at the $V\eta^\prime\gamma$ vertex.
Moreover, because of an anomalous gluon component, the structure function of
 the $\eta^\prime$ meson
is different from the one for  mesons which are
built up from quark-antiquark configurations. In a meson exchange 
model the intrinsic structure of the $\eta^\prime$ meson enters effectively
through the 
form factors at the $\rho\eta^\prime\gamma$ and $\omega\eta^\prime\gamma$ 
vertices, which do not necessarily have to be of the standard monopole or dipole 
type~\cite{Zhang1,Zhang2,Li,Borasoy1,Borasoy2}. In principle, the form 
factor at large $|t|$ is given by the meson structure function at short 
distances and in case of $\eta^\prime$ meson might be defined 
through its gluon component.

For our calculations we adopt an exponential form factor
\begin{equation}
{\tilde F}_{V\eta^\prime \gamma}(t)=\exp({\tilde \Lambda}_{V\eta^\prime \gamma} \, t),
\label{ff2}
\end{equation}
with ${\tilde \Lambda}_{V\eta^\prime \gamma}{=}1.2$~GeV$^{-2}$ being our best
choice value obtained by adjusting  to the $\eta^\prime$ meson photoproduction data.
Its behavior as function of the four momentum squared, $-t$, is shown by the
solid line in Fig.~\ref{prim6}(b). As comparison we show in the same figure 
as dashed line the monopole form factor of Eq.~(\ref{ff1}) with the 
standard~\cite{Elster}  cut-off parameter $\Lambda_{VNN}$=1~GeV.
For this comparison, the exponential form factor is renormalized 
to $F_{VNN}(t){=}1$ at $t{=}m_\rho^2$,
where $m_\rho$ being the $\rho$ meson mass. While the exponential and monopole 
form factors are almost identical for small $|t|$, i.e. $|t|{<}0.8$~GeV$^2$, 
the difference at
large --$t$ is quite obvious. 

Finally, the four Lorentz and gauge invariant  amplitudes~\cite{Chew} for the 
reaction $\gamma{p}{\to}\eta^{\prime}p$ due to the vector meson exchanges 
are given as
\begin{eqnarray}
A_1 &=& \frac{e g_{V\eta^\prime\gamma}}{m_{\eta^\prime}} \,
\frac{g_T}{2m_N} \, \frac{t}{t-m_V^2} \nonumber \\
A_1 + tA_2& =& 0 \nonumber \\
 A_3&=&0 \nonumber \\
A_4 &=& - \frac{e g_{V\eta^\prime\gamma} g_V }{m_{\eta^\prime}} \,
\frac{1}{t-m_V^2}.
\end{eqnarray}
It is understood that each amplitude contains in addition the 
product of the form factors at the 
$VNN$ and $V\eta^\prime\gamma$ vertices.

The differential cross section 
for the photoproduction of $\eta^\prime$ mesons
due to the vector meson exchange is obtained from
\begin{equation}
\frac{d\sigma}{dt} = \frac{F_{VNN}^2(t) F_{V\eta^\prime\gamma}^2(t)}
{16 \pi (s-m_N^2)^2 }\,
\sum_{i,j=1}^4 \!  Q_{i,j} \sum_{\rho , \omega} \!\!  A_i 
\sum_{\rho ,\omega}\!  A_j^\ast,
\label{xvect}
\end{equation}
where $Q_{ij}$ are the well known~\cite{Amaldi,Williams} hermitian matrix 
elements given in terms of Mandelstam invariants.


In Figs.~\ref{prim5a} and \ref{prim7a} we display the differential cross sections for 
the reaction $\gamma{p}{\to}\eta^\prime{p}$  
measured by the SAPHIR collaboration at ELSA~\cite{SAPHIR1,SAPHIR2}. We choose to 
plot the cross section
$d\sigma{/}dt$  instead of the angular spectra~\cite{SAPHIR1,SAPHIR2},
since in this case the slope with respect to $t$ explicitly indicates 
the strength of the 
form factor. Thus, the $t$ scale provides an appropriate representation of the 
data collected at large four momentum transfer squared.  


The solid lines in Figs.~\ref{prim5a} and \ref{prim7a}
represents our results obtained with an exponential form factor 
${\tilde F}_{V\eta^\prime \gamma}(t)$
at the $\omega\eta^\prime\gamma$ and $\rho\eta^\prime\gamma$
vertices with ${\tilde \Lambda}_{V\eta^\prime \gamma}{=}$1.2~GeV$^{-2}$,
and they reproduce 
for photon energies $E_\gamma{\ge}1.9$~GeV quite reasonably.

As a motivation of the form factor at the $\rho\eta^\prime\gamma$ and 
$\omega\eta^\prime\gamma$ vertices we analyze the experimental differential
cross sections in a parameterized form as
\begin{equation}
\frac{d\sigma}{dt} = A \exp(bt).
\end{equation}
The slope $b$ and the constant $A$ are shown in ig.~\ref{prim1} as
 function of the invariant collision energy $\sqrt{s}$. It should be noted
 that $A$ here does not represent the
magnitude of the differential cross section at $t{=}0$, as can be well 
understood from Fig.\ref{prim6}, but an extrapolated value.


Within  experimental uncertainties the slope $b$ does not depend on the energy
and can be well fitted by a constant value $b{=}1.8{\pm}0.2$~GeV$^{-2}$, 
as is shown by the solid line in Fig.\ref{prim1}(b). The $t$ dependence at low 
four momentum transfer squared beyond the resonance region is dominated by 
the vector exchange contribution. Therefore, the slope $b{=}1.8$~GeV$^{-2}$
at $\sqrt{s}{>}$2~GeV can be  linked to the  form factor at the 
$\rho\eta^\prime\gamma$ and $\omega\eta^\prime\gamma$ vertices. 

Possible contribution from other $\eta^\prime$ meson photoproduction 
mechanisms might change the slope $b$ at lower energies. If the $\eta^\prime$ 
meson couples to additional resonances, the exponential slope
of $t$ dependence may increase or decrease in the resonance region, 
depending whether an interference between the 
resonant and vector meson exchange  contributions is destructive or
constructive.  The size of the change of the slope with  the
photon energy will depend on the strength of the resonance coupling
to $\eta^\prime$ meson. However, as we found the uncertainties of 
the data are still too large to clarify this feature. 


Finally, since both, the slope of the $t$ dependence of $\eta^\prime$ meson 
photoproduction cross section and the constant $A$ do not depend on the 
energy at $\sqrt{s}{>}$2.0~GeV, 
we adjust the  cut off parameter 
${\tilde \Lambda}_{V\eta^\prime \gamma}$=1.2~GeV$^{-2}$ at these 
energies through~\cite{Sibirtsev2} the logarithmic derivative of the 
differential cross section. The solid line in Fig.~\ref{prim1}
shows the local slope at $t$=--0.6~GeV$^2$ evaluated  as
\begin{equation}
b(t,\sqrt{s})=\frac{d}{dt} \ln \left[ \frac{d\sigma}{dt} \right],
\label{slope}
\end{equation}
with the differential cross section for the reaction $\gamma{p}{\to}\eta^\prime{p}$ 
given by Eq.~(\ref{xvect}).


We also want to compare our calculation based on an exponential form
factor ${\tilde \Lambda}_{V\eta^\prime \gamma}$ to a calculation, where we 
employ a standard monopole form factor at the $\omega \eta^\prime \gamma$ and
$\rho \eta^\prime \gamma$ vertices. The dashed lines in Figs.~\ref{prim5a} and
\ref{prim7a} show such a calculation with a cut-off parameter
$\Lambda_{V\eta^\prime \gamma}{=}$1.5~GeV at the $\omega \eta^\prime \gamma$ and
$\rho \eta^\prime \gamma$ vertices. This value is identical to the one at the
$\omega NN$ and $\rho NN$ vertices. We see that a calculation with a monopole
form factor at the electro-magnetic vertices can not reproduce the $t$ dependence
of the $\eta^\prime$ meson photoproduction data especially at photon energies
above 1.7~GeV. It should not be surprising, that form factors as the vertices
with the vector mesons coupling to the nucleon are different from the ones where
the vector mesons couple to the photon, since the internal structure of those
vertices is different.

The comparison between the SAPHIR data~\cite{SAPHIR1,SAPHIR2} and our calculations
based on vector meson exchange as sole production mechanism leads  
to the following conclusions. 
First, a comparison of the calculations to the data 
for small photon energies,  $E_\gamma{\le}1.6$~GeV, shows some room for a possible 
contributions from  $s$ wave resonances. However, we do not
see much room for the contribution from the $p$ wave resonance, which must
introduce an additional $t$ dependence. Second, the differential cross section at 
$E_\gamma{=}2.1$~GeV seems to 
indicates some enhancement at maximal --$t$ or  small $|u|$, which can not
be explained by considering vector meson exchange alone. Indeed, this enhancement
may point to the importance of considering the nucleon exchange as part of the
production mechanism.


\section{Contributions from  Baryon Resonances }

The possibility  of studying excited baryons which couple to the $\eta^\prime$ meson
in the  in the reaction $\gamma{p}{\to}\eta^{\prime}p$ initiated considerable
theoretical activity~\cite{Zhang1,Zhang2,Li,Borasoy1,Borasoy2,Chiang}, in which  
contributions from a number of resonances were introduced.  For example,
the approach of
Refs.~\cite{Zhang1,Zhang2}, based on effective Lagrangians, 
considers nine baryonic resonances in different partial waves.
These are 
two $S_{11}$ resonances as $N^\ast(2030)$ and $N^\ast(2090)$, three $D_{13}$
resonances as $N^\ast(2055)$, $N^\ast(2080)$ and $N^\ast(2095)$, 
two $D_{15}$ resonances, $N^\ast(2080)$ and $N^\ast(2200)$ and one $F_{17}$
resonance, $N^\ast(1990)$. Contributions from only the $S_{11}(2090)$, 
$P_{11}(2100)$ and $P_{13}(1900)$ resonances
are discussed in the Regge model calculation 
of  Ref.~\cite{Chiang}. In contrast, the quark model 
calculation of  Ref.~\cite{Li} considers  only  a $S_{11}(1535)$ resonance.

Because of the large uncertainties in the experimental results and the 
very limited number of experimental points~\cite{SAPHIR1,SAPHIR2} presently 
available for the reaction  $\gamma{p}{\to}\eta^{\prime}p$, we prefer to 
introduce as few resonances into our calculation as possible. It is also clear, 
that large uncertainties in the selection of
the parameters of potential resonances, i.e. mass and width,  the coupling to the
$\eta^\prime$ meson as well as our lack of knowledge about relevant form factors at the
resonance vertices might provide quite a large freedom in describing the present data by
isobar contributions. Having this in mind, and 
considering our results from Figs.~\ref{prim5a} and \ref{prim7a}, we see that 
we mostly underpredict the present data close to the reaction threshold. 
Overall, we find that the data can be reasonably well described by a single production
mechanism, namely the exchange of vector mesons.
It is also clear that  this exchange of  vector mesons 
can be reggeized~\cite{Chiang}, providing it is applicable at the
very high energies. Thus, concluding from Fig.~\ref{prim1}  we only have
room for 
additional contributions at photon energies  $E_\gamma{<}1.7$~GeV, or at invariant 
collision energies below 2~GeV. Furthermore, since the slope $b$ of the 
$t$ dependence does not change at  $E_\gamma{<}1.7$~GeV, we also do not see
room for the contribution from any $p$ wave (or higher partial wave) resonances.  

A closer look at Fig.\ref{prim1}(a) reveals,  that this possible additional contribution
dominates already at the $\eta^\prime$ meson photoproduction threshold,
$\sqrt{s}{\simeq}1896$~MeV, decreases with increasing photon energy and  almost 
vanishes for $\sqrt{s}{\simeq}2010$~MeV. Since the differential cross section,
$d\sigma{/}dt$,  
given in Eq.(\ref{xvect}) does not contain a final phase space 
factor, the constants $A$ can be considered as being proportional to the 
squared averaged reaction matrix element, which has a maximum value at 
threshold and saturates with increasing the photon energy.  Thus we 
guess that  energies with $1896{\le}\sqrt{s}{\le}2010$~MeV  (see Fig.\ref{prim6}(a))
are influenced by the tale of a resonance which has a  pole at or below the 
$\eta^\prime$ meson photoproduction threshold. 

Since there are little constraints on resonances coupled to the $\eta^\prime$ meson, 
different choices of $s$ wave resonances having masses 
below  1896~MeV  are in principle possible.
In general,
the interaction Lagrangians for a  $N(\frac{1}{2}^\pm)$ nucleon  resonances 
 are given 
as~\cite{Levy,Olsson1,Olsson2,Benmerrouche1,Benmerrouche2}
\begin{eqnarray}
{\cal L}_{\eta^\prime N R} =-i g_{\eta^\prime NR} \,
{\bar R} \Gamma N \eta^\prime + {\mbox h.c.},
\nonumber \\
{\cal L}_{\gamma N R}
= e {\bar R}\, \frac{\kappa^S_R{+}\kappa_R^V\tau_3}{2(m_N{+}m_R)}
{\tilde \Gamma} \sigma_{\mu\nu} N F^{\mu\nu}{+}{\mbox h.c.},
\end{eqnarray}
where $m_R$ denotes the resonance mass, $R$ the field operator of the resonance,
and  $\kappa_R^S$ and $\kappa_R^V$ are the
isoscalar and isovector transition magnetic couplings. The vertex functions
for nucleon resonances $R$ with negative parity are $\Gamma$=1 and 
${\tilde\Gamma}{=}\gamma_5$, while resonances with positive parity have 
$\Gamma{=}\gamma_5$ and ${\tilde\Gamma}{=}$1. 


Furthermore, for a proton target $\kappa_R^p{=}\kappa_R^S{+}\kappa_R^V$
and the relation between the magnetic coupling and the helicity amplitude
$A_{1/2}^p$ is given by~\cite{Benmerrouche2,Mitra,Copley} 
\begin{equation}
|A_{1/2}^p|^2 = \left( \frac{e\kappa_R^p}{m_N+m_R}\right)^2
\frac{m_R^2-m_N^2}{2m_N}.
\end{equation}
For a neutron target one has $\kappa_R^n{=}\kappa_R^S{-}\kappa_R^V$, and 
the helicity amplitudes $R{\to}p\gamma$ and $R{\to}n\gamma$ are given 
by the Particle Data Group~\cite{PDG}. We take $\kappa_R$=2.3. 
The couplings of the $\eta^\prime$ meson to baryon resonances, 
$g_{\eta^\prime NR}$, are generally unknown and can only be adjusted to data.
In our calculation a value $g_{\eta' RN}$=3.4 is adopted.

Finally, the invariant amplitudes for  $s-$ and $u$-channel  
nucleon resonance contributions can be written  as
\begin{eqnarray}
A_1 & =& \pm \frac{e g_{\eta^\prime NR} \kappa_R 
(m_N\pm m_R)}{m_N+m_R}\left[ \frac{1}{s{-}m_R^2}{+}\frac{1}{u{-}m_R^2} \right],
\nonumber \\
A_2&=&0  \nonumber \\
A_3& =& \pm \frac{e g_{\eta^\prime NR} \kappa_R}
{m_N+m_R}\left[ \frac{1}{s{-}m_R^2}{-}\frac{1}{u{-}m_R^2} \right],
\nonumber \\
A_3 & = & \pm \frac{e g_{\eta^\prime NR} \kappa_R}
{m_N+m_R}\left[ \frac{1}{s{-}m_R^2}{+}\frac{1}{u{-}m_R^2} \right],
\end{eqnarray}
where the plus sign is assigned to the negative parity excited state, while  the
minus sign corresponds to positive parity excited state. The coupling $\kappa_R$ stands 
for the  magnetic coupling for a proton or a  neutron target.

Furthermore, we include form factors at the interaction vertices, because 
an excited, intermediate baryon resonance is off shell. In our calculations we employ an 
overall formfactor with the  general form~\cite{Davidson}
\begin{equation} 
F(u,s)=F(u)+F(s)-F(u)F(s).
\label{david}
\end{equation}
The individual formfactors in the $u$ and $s$ channels are given by
\begin{equation}
F(r)=\frac{\Lambda_R^4}{\Lambda_R^4-(r-m_R^2)^2}.
\label{indi}
\end{equation}
Here  $r{=}s$ or $u$, and the cut off parameter $\Lambda_R$ is adjusted to the data. 
In our calculations we use $\Lambda_R$=1~GeV, and take the same values
 for both, the  $u$ and the $s$ channels.

To study the influence of  s-wave resonances on the behavior of the differential
cross section close to the production threshold, we prefer to select only the 
well known $S_{11}(1535)$ resonance and investigate,
whether its contribution is sufficient to account for the
difference between our calculation based on vector meson exchange alone
and the data for photon energies $E_\gamma{<}$1.7~GeV. 


Our calculations including the contribution of the resonance $S_{11}(1535)$
as well as the vector meson exchange
is shown as solid line in Fig.~\ref{prim5b} in comparison to the data for 
photon energies $1.5{\le}E_\gamma{\le}2.1$~GeV. The dashed line represents the
calculation with the vector  meson exchange alone. 
The resonant contribution dominates  around
the $\eta^\prime$ meson photoproduction threshold and vanishes with
increasing the photon energy.

Concluding this section we want to emphasize that our calculations
do not  prove that resonant contributions to the $\eta^\prime$
meson photoproduction are not necessarily only those from the 
$S_{11}(1535)$ resonance.
Large uncertainties in selection of the coupling constants $g_{\eta^\prime{NR}}$ to 
resonances 
as well as in cutoff parameters $\Lambda_R$ allow sufficient freedom to consider
more resonant contributions. However, in this work, we did not prefer to include
any other resonance than $S_{11}(1535)$.

\section{Contribution of the Nucleon Exchange}

Since the mass of the $\eta^\prime$ meson is close to the nucleon mass,
the range of the squared four momentum $u$ available in the reaction
$\gamma{p}{\to}\eta^{\prime}p$  is almost the same as
the range of the variable $t$ as is indicated in Fig.~\ref{prim6}. 
As $-u$ approaches its minimal value, the contribution from the
nucleon exchange increases due to the $u$ dependence of 
the squared nucleon propagator,  $(u{-}m_N^2)^{-2}$. 

To map out the $u$ dependence of the differential cross section, 
it is worthwhile to perform the measurements at sufficiently large 
photon energies, which provide  access to a wide range of $u$.
In addition, at large energies the separation between the $t$ and 
the $u$ channel becomes more pronounced. At the same time, the 
$s$ channel contribution from the nucleon exchange is small
because of the $s$ dependence of the squared nucleon propagator,
$(s{-}m_N^2)^{-2}$. 

Obviously,  baryon resonances may also contribute  at
small $|u|$. However, they contribute at the same time  to the $s$
channel and fill the gap between small --$t$ and small --$u$.
Though it may not be possible to evaluate  resonance properties unambiguously
from the data, the overall resonant contribution
can be fitted, as we illustrated in the previous section.
Thus the resonant background to the $u$ channel nucleon exchange
contribution can be well estimated and subtracted. 
 
In general, the contribution from the nucleon exchange current can be
measured at small $u$ or at backward angles in meson
photoproduction reaction. Obviously, the size of the contribution will depend on the
strength of the coupling of the $\eta^\prime$ meson to the nucleon. 
In fact, such an increase of the differential
cross section at small $u$ was detected in the photoproduction
of $\pi$ and $\omega$ mesons~\cite{Anderson,Clift}. Thus, it may also be observed in
the photoproduction of $\eta^\prime$ mesons.

The effective Langrangians for  the $\gamma{NN}$ and $\eta^\prime{NN}$ interaction
can be written as~\cite{Gross}
\begin{eqnarray}
{\cal L}_{\gamma NN} & = & {-}e{\bar N}\left( \gamma_\mu \frac{1{+}\tau_3}{2}
A^\mu {-}\frac{\kappa^S{+}\kappa^V\tau_3}{4m_N}  \sigma_{\mu\nu}
F^{\mu\nu} \right)N \nonumber \\
{\cal L}_{\eta^\prime{NN}} & = & -i g_{\eta^\prime{NN}} {\bar N}\gamma_5
N \eta^\prime. \hspace*{5mm}
\end{eqnarray}
Here $\kappa^S$ and $\kappa^V$ are the isoscalar and isovector
anomalous magnetic moments of the nucleon,
$\kappa_p{=}\kappa^S{+}\kappa^V{=}1.79$  and
while $\kappa_n{=}\kappa^S{-}\kappa^V{=}{-}1.91$ stand for the proton
and neutron, respectively.

The four invariant amplitudes for the $\eta^\prime$ meson photoproduction
due to the nucleon exchange current in $s$ and $u$ channels are
given as 
\begin{eqnarray}
A_1 & =& e_Neg_{\eta^\prime NN}\left[\frac{1}{s-m_N^2}{+}
\frac{1}{u-m_N^2} \right], \nonumber \\
A_2 & = & \frac{2e_N e g_{\eta^\prime NN}}{(s-m_N^2)(u-m_N^2)},
\nonumber \\
A_3 & = & -\frac{e g_{\eta^\prime NN} \kappa_N}{2m_N}
\left[\frac{1}{s-m_N^2}{-}
\frac{1}{u-m_N^2} \right], \nonumber \\
A_4 & = & -\frac{e g_{\eta^\prime NN} \kappa_N}{2m_N}
\left[\frac{1}{s-m_N^2}{+}
\frac{1}{u-m_N^2} \right],
\end{eqnarray}
where $e_N{=}0$ for a neutron and $e_N{=}1$ for a proton target.
In principle, we should consider formfactors at the interaction vertices, since the 
nucleons are off shell in the intermediate state. However, the 
introduction of a form factor violates gauge invariance. 
To prevent this, we use  the form factor given by Eqs.(\ref{david},\ref{indi})
with a cutoff parameter $\Lambda_N$=800~MeV.

As we discussed in the introduction, the coupling constant $\eta^\prime{NN}$  is
unknown and we suggest the possibility to obtain information and constraints on it 
from the $\eta^\prime$ meson
photoproduction data at small $|u|$. However, within accuracy of the present
data shown in Fig.~\ref{prim8},  it is quite speculative to discuss 
additional contributions and any enhancement of the
differential cross section for the reaction  $\gamma{p}{\to}\eta^{\prime}p$ 
at small $|u|$. Thus, for the present calculations 
of the nucleon exchange contribution we only estimate the size of the
$\eta^\prime{NN}$ coupling constant.


Because of the $SU(3)$ breaking the pseudoscalar octet mixes with the 
corresponding pseudoscalar singlet to produce the physical $\eta$ and 
$\eta^\prime$ meson states. Taking the $\eta$-$\eta^\prime$ mixing angle
$\theta{\simeq}{-}20^o$ and the ratio of the singlet to octet constants
${\simeq}\sqrt{2}$~\cite{Koniuk}, the $\eta{NN}$ and $\eta^\prime{NN}$
coupling constants can be related as
\begin{equation}
\frac{g_{\eta^\prime{NN}}}{g_{\eta{NN}}} =\frac{ \sin\theta + \sqrt{2}\cos\theta}
{cos\theta -\sqrt{2}\sin\theta}.
\end{equation}

A theoretical analysis~\cite{Benmerrouche1,Benmerrouche2} of the $\eta$ meson 
photoproduction data gives $2.7{\le}g_{\eta{NN}}{\le}8.9$ and
consequently $1.9{\le}g_{\eta^\prime{NN}}{\le}6.1$. Estimates based on  $SU(6)$ 
\cite{Tornqvist} give  $g_{\eta^\prime{NN}}{=}6.5$.  

The $\eta^\prime{NN}$ coupling constant was also extracted from
experimental observables. An old analysis~\cite{Dumbrajs} based on $NN$ potentials
gives $g_{\eta{NN}}{=}6.8$ and $g_{\eta^\prime{NN}}{=}7.3$, while 
nucleon-nucleon forward dispersion relations~\cite{Grein}
yield $g_{\eta{NN}}^2{+}g^2_{\eta^\prime{NN}}{\le}12$. 
Calculations of the decay $\eta^\prime{\to}\gamma\gamma$  with baryon 
triangle contributions~\cite{Bagchi} give $g_{\eta^\prime{NN}}{=}6.3$.
Although, the  $\eta^\prime{NN}$ constant extracted from different
experiments is in a reasonable agreement with  the $SU(3)$ estimates, 
the exact value is still considered as being model dependent, since the gluonic
component of the $\eta^\prime$ meson was neglected. For example, in $NN$
scattering the gluon-nucleon interaction generates a contact term
and may reduce the extracted value of the  $\eta^\prime{NN}$ coupling 
constant.

Neglecting the gluon-nucleon coupling $g_{GNN}$ in Eq.(\ref{GD}) one
can estimate the  $\eta^\prime{NN}$ coupling constant as
$g_{\eta^\prime{NN}}{\simeq}2.2$. This value can be considered as  quite small,
as we discussed in the introduction. On the other hand, the OZI
predictions~\cite{Ellis1,Ellis2} for the axial charge of the nucleon, 
neglecting the  polarized strange quark and gluon contributions to the
internal spin structure of the nucleon, gives $g_A^0{\simeq}0.6$. Attributing this
value to the $\eta^\prime{NN}$ coupling constant one can estimate 
$g_{\eta^\prime{NN}}{\simeq}4.6$. The more careful analysis~\cite{Cheng} 
based on the Goldberger-Treiman relation for the axial charge gives 
$g_{\eta^\prime{NN}}{=}3.4$

For our calculations we adjust the coupling constant ${\eta^\prime{NN}}$ within
the above listed boundaries to the data and use  $g_{\eta^\prime{NN}}{=}2.9$.
The solid lines in Fig.~\ref{prim8} shows our calculations for the
$\eta^\prime$ photoproduction cross section based on the
contribution from the vector meson exchange,  the $S_{11}(1535)$ resonance and the 
nucleon exchange current.
As we can see, the quality of the presently available data, especially at large $|t|$,
is such, that it is not useful to estimate any uncertainty of value of
$g_{\eta^\prime{NN}}$.
 However our calculations presented in Fig.~\ref{prim8} clearly illustrate that 
the differential cross section for $\eta^\prime$ meson photoproduction at small
$|u|$ or at large scattering angles can be used for a determination of the
${\eta^\prime{NN}}$ coupling constant.

\section{Summary and Conclusions}

Our goal was to describe the reaction ${\gamma}p{\to}{\eta^\prime}p$ from
threshold to  photon energies $E_\gamma$=2.4~GeV.
First we studied the contribution from vector meson exchange in the $t$
channel to the $\eta^\prime$ meson photoproduction cross section.
With this, the presently available data from the  
SAPHIR collaboration at ELSA~\cite{SAPHIR1,SAPHIR2}
can be well reproduced for  photon energies $E_\gamma{\ge}1.8$~GeV to 2.4~GeV. 

The discrepancy between our calculation based on vector meson exchange alone and the 
data at small
photon energies, $1.5{\le}E_\gamma{<}1.7$~GeV, may be attributed to 
contributions of resonances. An  analysis of the data 
in terms of the energy dependence
of the extrapolated forward  differential cross section $d\sigma{/}dt$
supports the assumption that this additional  contribution results from
a resonance with a  mass below the  $\eta^\prime$ meson production
threshold. Thus we include the well established 
$S_{11}(1535)$ nucleon
resonance into our calculation. Finally we can
 describe  the available data for  $1.5{\le}E_\gamma{\le}2.4$~GeV quite well. 
However we note, that with the presently available data we can not conclude that
the $S_{11}(1535)$ resonance is the only one that
contributes to the reaction $\gamma{p}{\to}\eta^\prime{p}$. 

As last step we investigated the contribution from the nucleon exchange to the
reaction  $\gamma{p}{\to}\eta^\prime{p}$. The aim is to see whether this contribution
can be detected experimentally. Our calculations show 
that the nucleon exchange contribution
dominates at small $|u|$ for photon energies $E_\gamma{\ge}1.8$~GeV.
Our speculation is that an experimental measurement under the above kinematical
conditions
can be used for a better evaluation of the contribution from the
nucleon exchange, which in turn allows to  
obtain  the ${\eta^\prime{NN}}$ coupling
constant. Because of the large uncertainties of available estimates
for the ${\eta^\prime{NN}}$ coupling constant and its importance for the
clarification of  proton spin crisis
problem~\cite{Veneziano1,Veneziano2,Hatsuda,Efremov,Chao},
a precise measurement of the  $\eta^\prime$ meson photoproduction at large angles
is crucial.

\begin{acknowledgments}
This work was performed in part under the auspices of the
U.~S. Department of Energy under contract No. DE-FG02-93ER40756
with the Ohio University.
The authors appreciate very useful discussions and communications with M.~Dugger, 
J.~Ernst, E.~Pasyuk and B.G.~Ritchie. A.S would like to acknowledge the 
warm hospitality of the Department of Physics and Astronomy at
 Ohio University during his visit.
\end{acknowledgments}


\newpage

\begin{figure}
\caption{
Diagrams contributing to the photoproduction of the $\eta^\prime$ meson from a
nucleon: (a) denotes the vector meson exchange of $\omega$ and $\rho$, (b) the
u-channel contribution of the nucleon N (or resonance R) exchange, 
and (c) stands for the corresponding s-channel contribution.}
\label{diagram1}
\end{figure}

\begin{figure}
\caption{
 The shaded area in (a) shows the range of $t$ and $s$ accecible in
reaction $\gamma N  \to \eta^{\prime}N$ at the photon energy $E_\gamma$ given on
the upper axis.
In (b) the behavior of 
the monopole (dashed) and exponential (solid line) form factors are shown
as  function of the four momentum transfer squared --$t$. The
form factors are shown for $\Lambda_V$=1.0~GeV and 
${\tilde \Lambda}_{V\eta^\prime \gamma}$=1.2~GeV$^{-2}$ 
and normalised $F_{V\eta^\prime \gamma}(t){=}1$ at $t{=}m_\rho^2$.}
\label{prim6}
\end{figure}

\begin{figure}
\caption{
The differential cross section for the reaction $\gamma p  \to \eta^{\prime}p$ 
as a function of the four momentum transfer squared $t$. The squares represent the
first
SAPHIR results given in Ref.~\protect\cite{SAPHIR1} with absolute normalization. The
circles stand for new SAPHIR data reported in Ref.~\protect\cite{SAPHIR2} in relative
normalization,
which we multiply by  a factor of 2.9 at all given
photon energies $E_\gamma$.  The lines represent our calculations, for the solid
line the 
exponential form factor ${\tilde F}_{V\eta^\prime \gamma}(t)$ of
 Eq.~(\protect\ref{ff2}) at the vertices
$\gamma\rho\eta^\prime$ and $\gamma\omega\eta^\prime$  is employed, for the
dashed line the 
 monopole form factor $F_{V\eta^\prime \gamma}(t)$ of Eq.~(\protect\ref{ff1})
is used.}
\label{prim5a}
\end{figure}

\begin{figure}
\caption{
Same as in Fig.~\ref{prim5a} for different photon energies
$E_\gamma$.}
\label{prim7a}
\end{figure}

\begin{figure}
\caption{
The extrapolated forward $\eta^\prime$ meson photoproduction cross
section (a)  and the slope $b$ of the $t$ dependence (b) as a function of
the invariant
collision energy. The squares show the first SAPHIR results
from Ref.~\protect\cite{SAPHIR1}
with absolute normalization, while the circles are new SAPHIR data
reported in Ref.~\protect\cite{SAPHIR2} in relative normalization and multiplied by 
a factor of 2.9. 
The arrows indicate the reaction threshold.  The dashed line in (b)
shows the fit by a constant value $b{=}1.8{\pm}0.2$~GeV$^{-2}$, while the
solid line uses the local slope at $t$=--0.6~GeV$^2$ resulting from our
calculations with Eq.~(\protect\ref{slope}).}
\label{prim1}
\vspace*{-0.1mm}
\end{figure}

\begin{figure}
\caption{
The differential cross section for the reaction $\gamma{p}{\to}\eta^{\prime}p$ 
as  function of four momentum transfer squared $t$. The squares show the
first
SAPHIR results given in Ref.~\protect\cite{SAPHIR1} with absolute normalization. The
circles are new SAPHIR data reported in Ref.~\protect\cite{SAPHIR2} in relative
normalization,
which we multiply by  a factor of 2.9 at all given
photon energies $E_\gamma$.  The  dashed line represents our calculation with
the vector meson exchange alone as production mechanism (corresponding to the
solid line in Figs.~\protect\ref{prim5a} and \protect\ref{prim1}). The  solid
line includes in addition the contribution from
the $S_{11}(1535)$ resonance.}
\label{prim5b}
\end{figure}

\begin{figure}
\caption{
The differential cross sections for the $\eta^{\prime}$ meson photoproduction 
as a function of four momentum transfer squared $t$ at different photon
energies $E_\gamma$. The squares show the first
SAPHIR results given~\protect\cite{SAPHIR1} with absolute normalization. The
circles are new SAPHIR data reported~\protect\cite{SAPHIR2} in relative
normalization,
which we multiply by  a factor of 2.9 at all given
photon energies $E_\gamma$. The solid line represents our calculation including
vector meson exchange, the contribution from the $S_{11}(1535)$ resonance as well
as the contribution from the nucleon exchange. For the calculation shown as dashed
line, the contribution from the nucleon exchange was omitted.}  
\label{prim8}
\end{figure}



\newpage

\begin{figure}
\begin{center}
\psfig{file=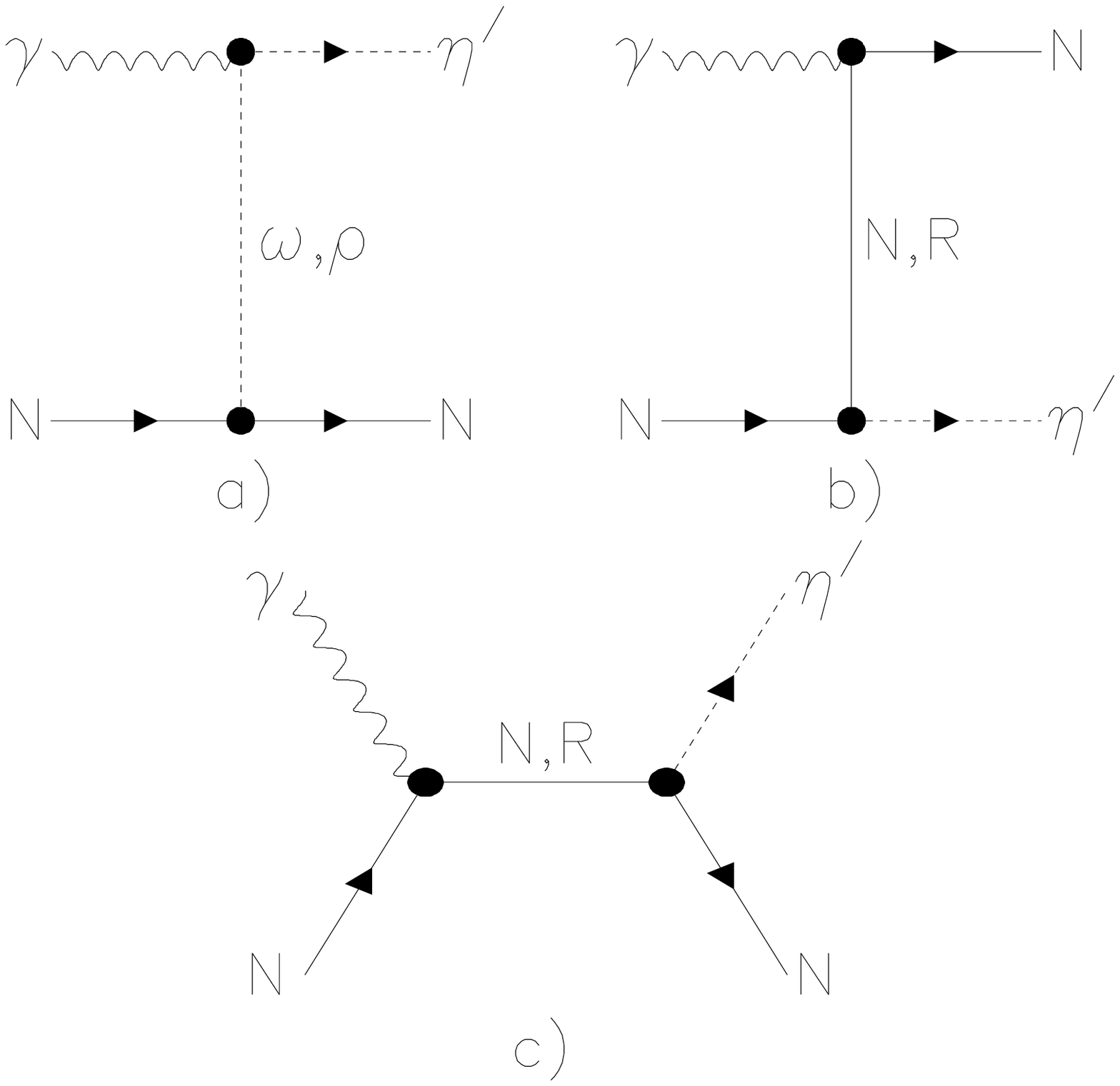,width=8cm,height=8cm}
\center{FIG. 1}
\end{center}
\end{figure}

\begin{figure}
\begin{center}
\psfig{file=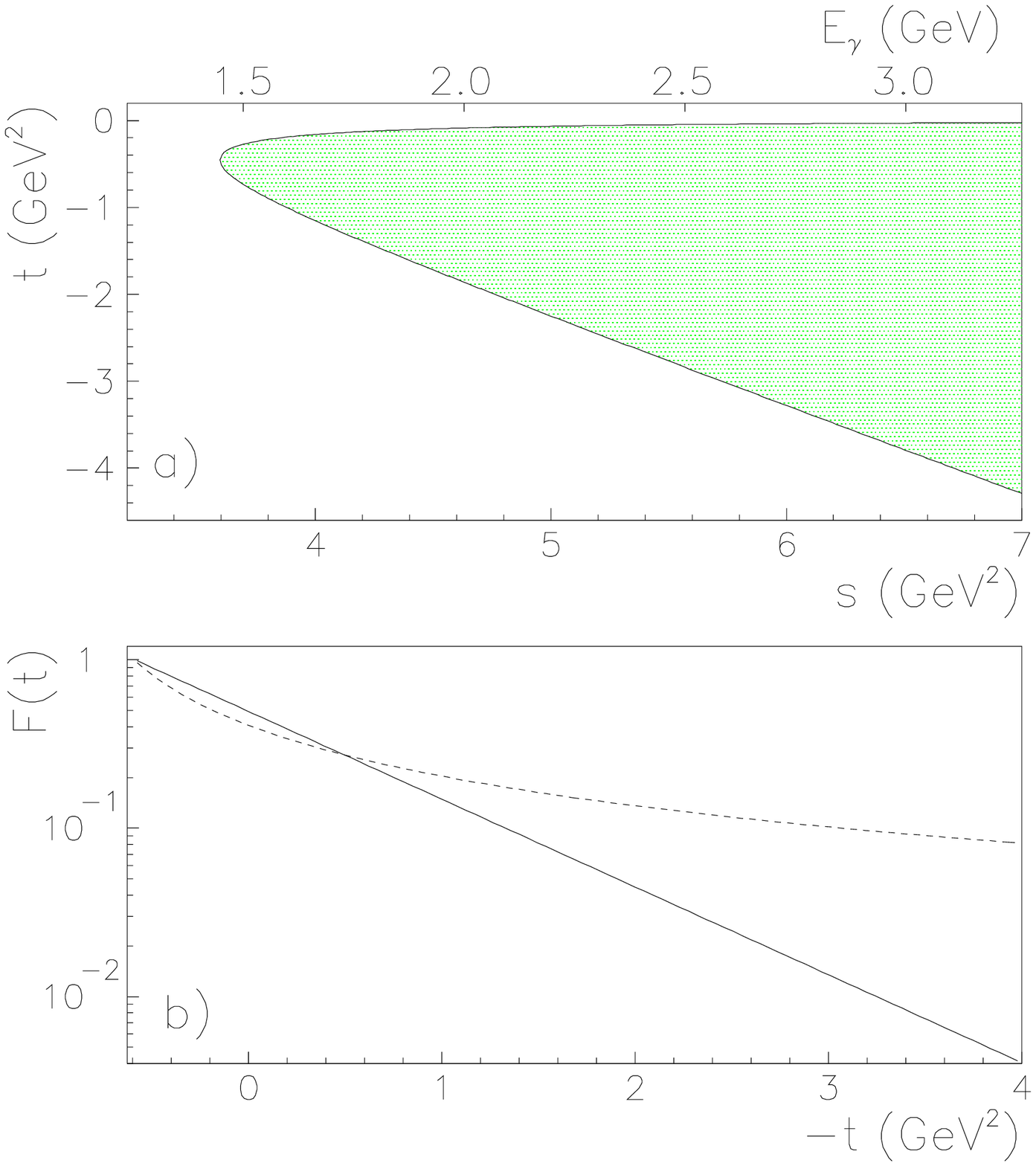,width=9.5cm,height=9.cm}
\center{FIG. 2}
\end{center}
\end{figure}

\begin{figure}
\begin{center}
\psfig{file=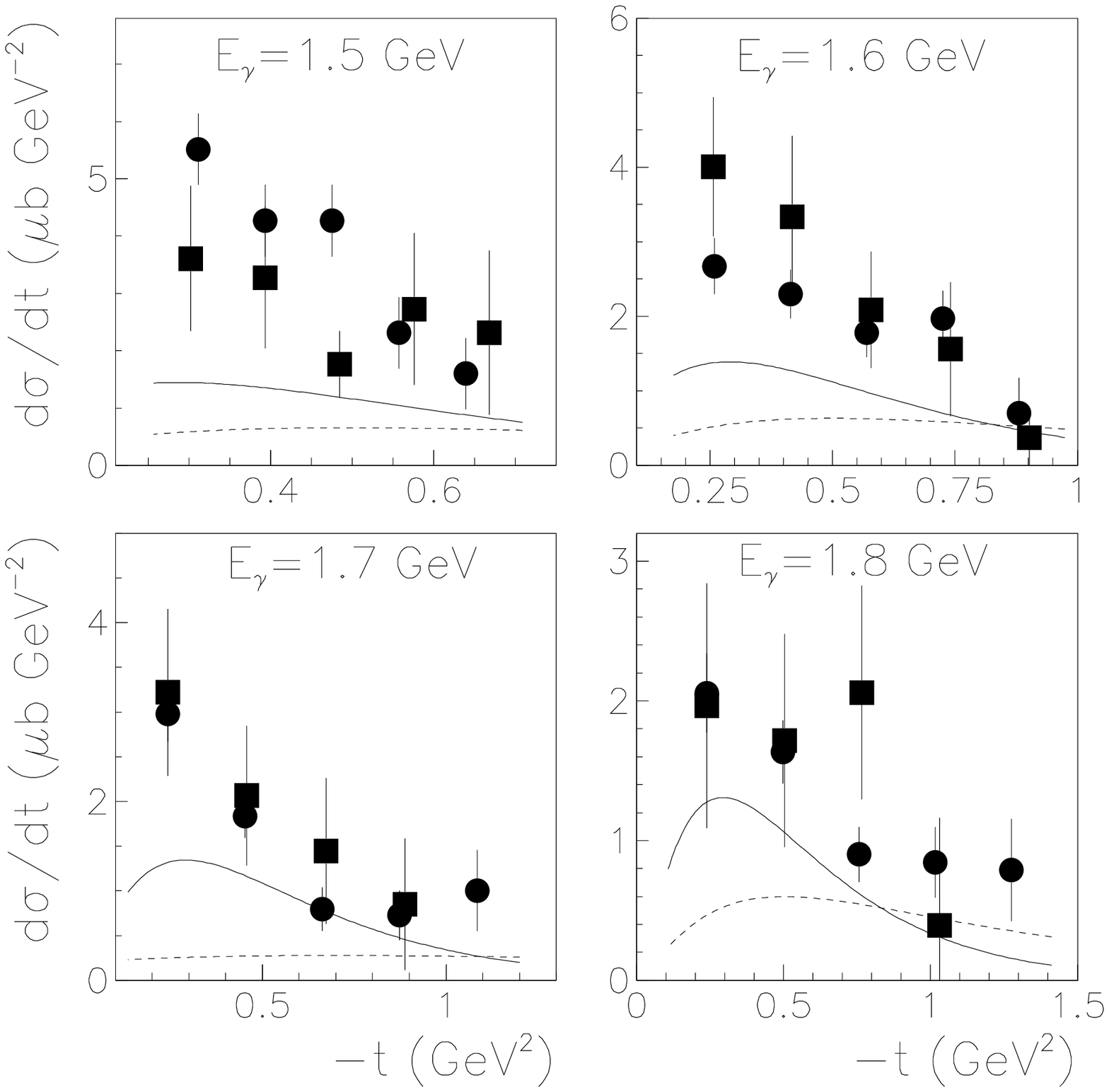,width=11.6cm,height=10.6cm}
\vspace{-4mm}
\center{FIG. 3}
\end{center}
\end{figure}

\vspace{-9mm} 

\begin{figure}
\begin{center}
\psfig{file=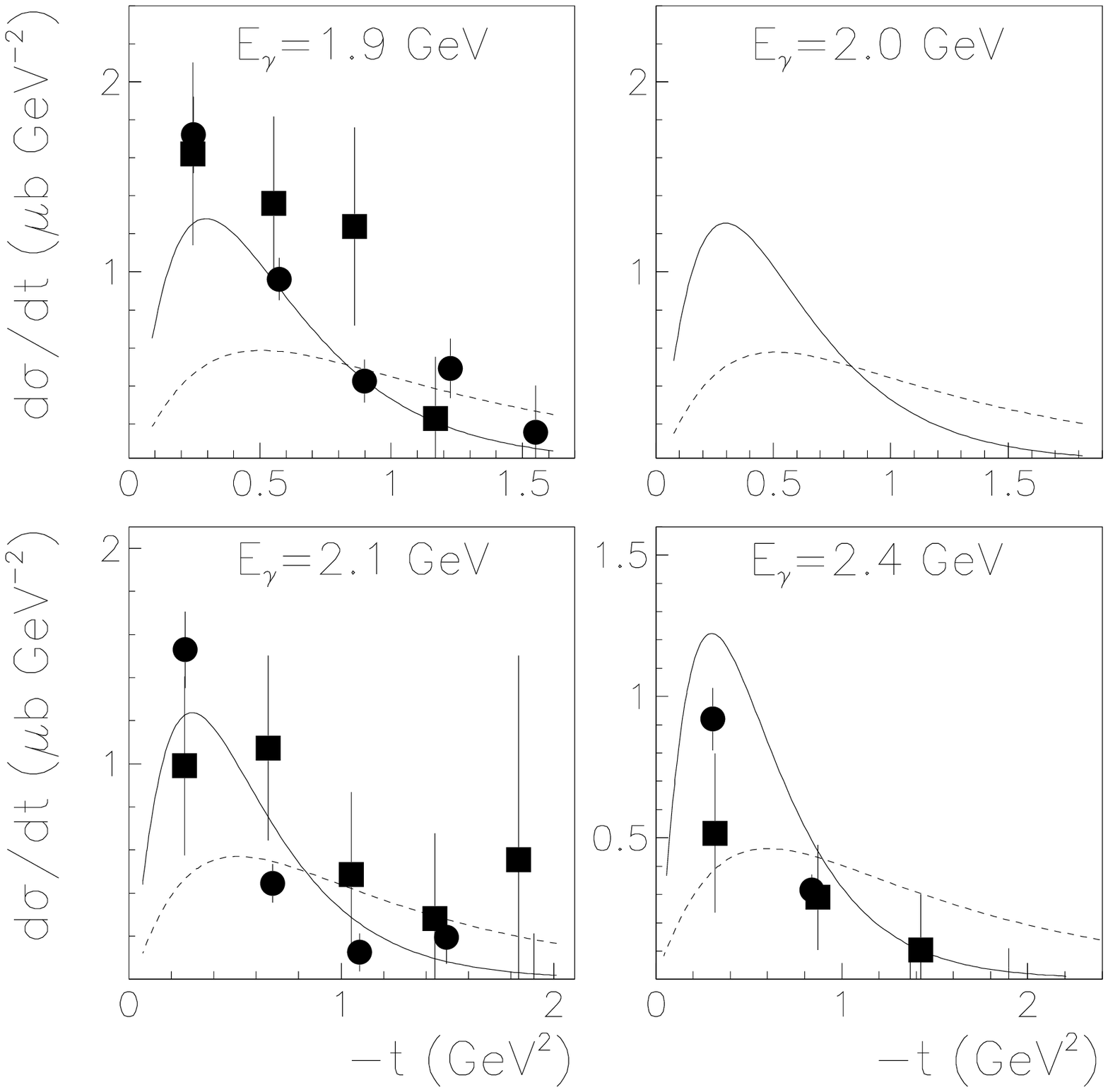,width=11.6cm,height=10.6cm}
\vspace{-2mm}
\center{FIG. 4}
\end{center}
\end{figure}

\begin{figure}
\begin{center}
\psfig{file=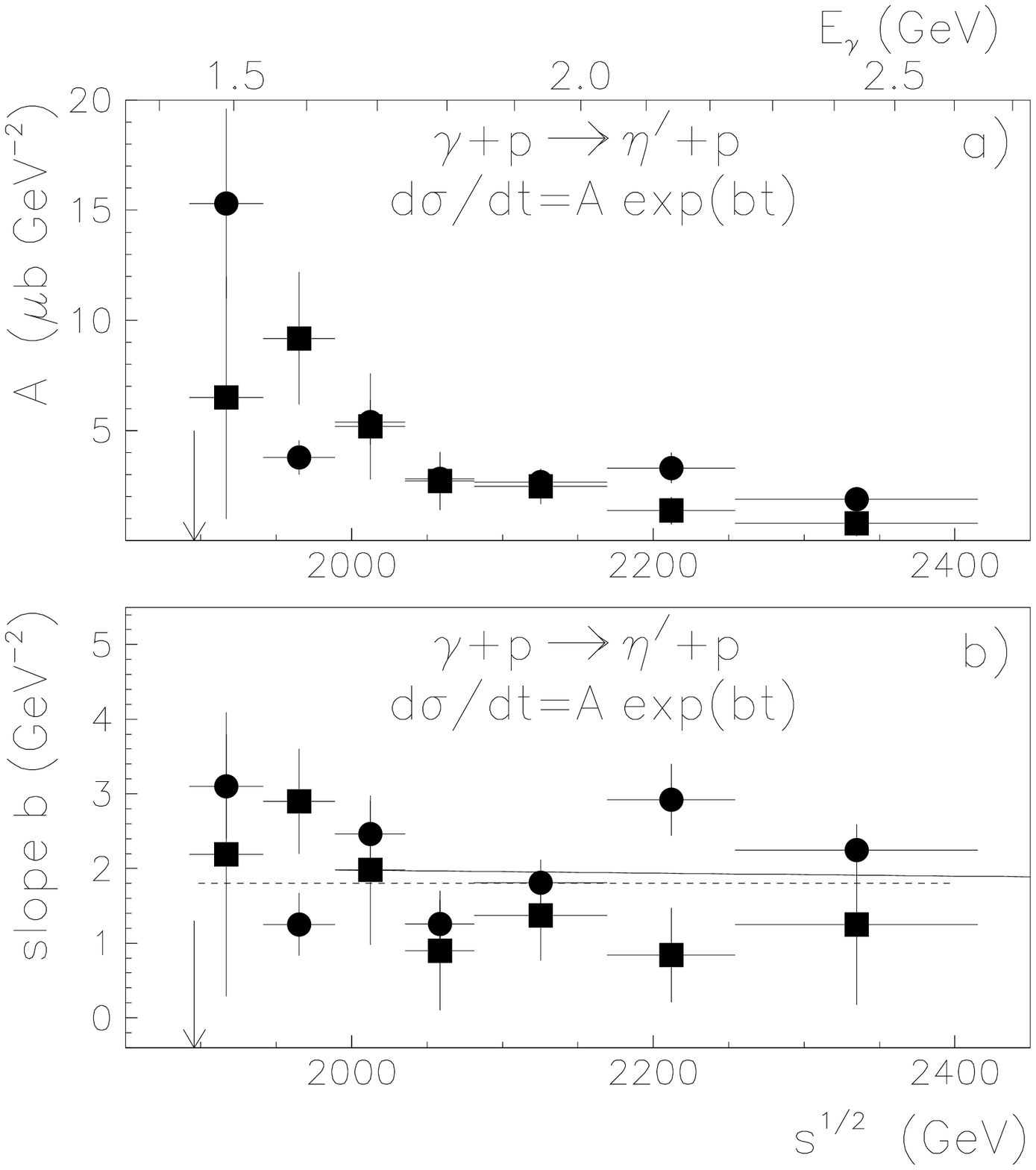,width=9.2cm,height=9.cm}
\vspace{-2mm}
\center{FIG. 5}
\end{center}
\end{figure}

\vspace{-5mm} 

\begin{figure}
\begin{center}
\psfig{file=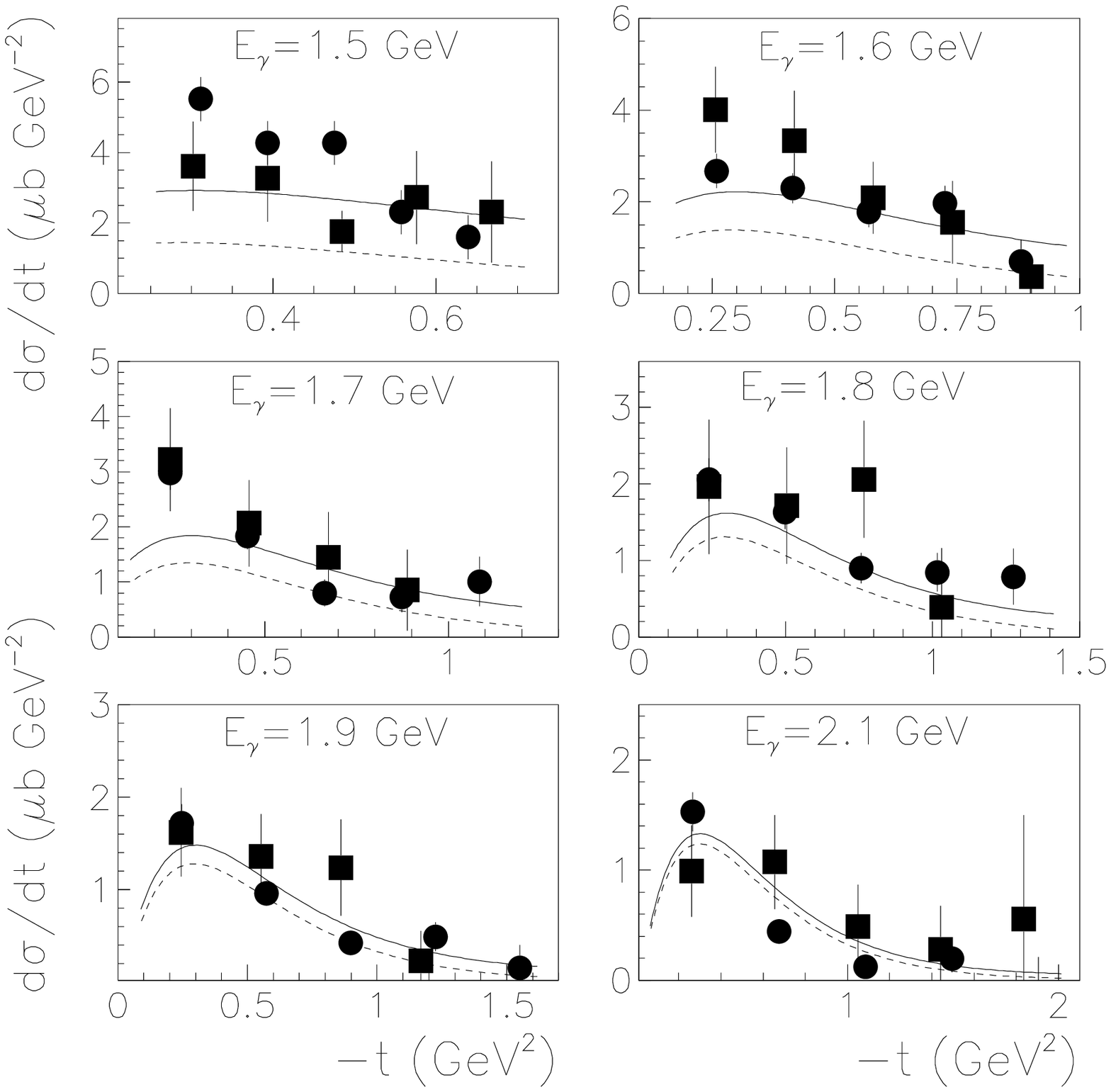,width=11.3cm,height=10.5cm}
\vspace{-2mm}
\center{FIG. 6}
\end{center}
\end{figure}


\begin{figure}
\begin{center}
\psfig{file=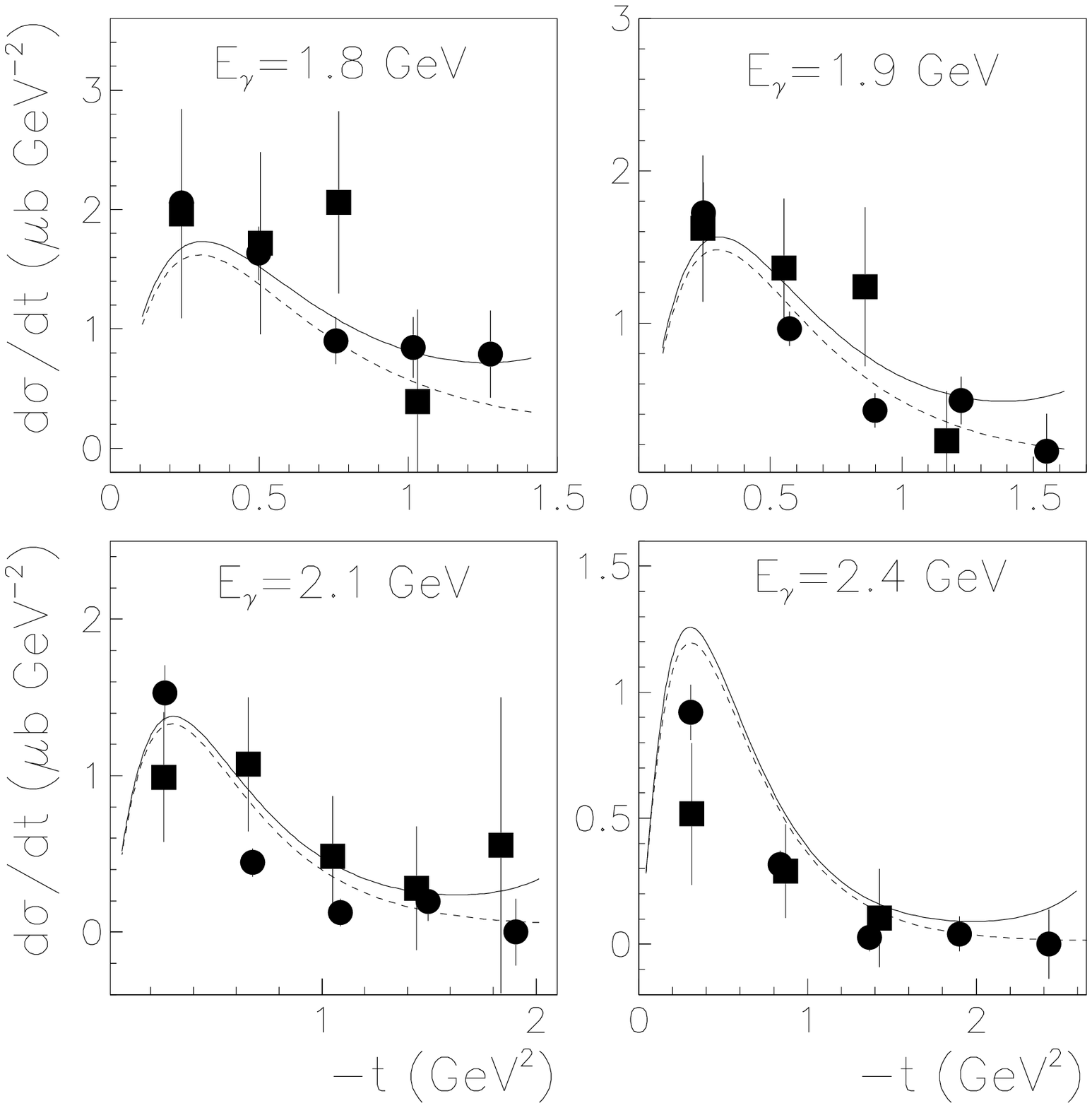,width=11.3cm,height=10.0cm}
\center{FIG. 7}
\end{center}
\end{figure}

\end{document}